\documentclass[11pt,a4paper]{article}
\usepackage[hyperref]{emnlp2018}
\usepackage{times}
\usepackage{latexsym}

\usepackage{url}

\usepackage{graphicx}

\aclfinalcopy 

\title{Uni-DUE Student Team: Tackling fact checking through decomposable attention neural network}

\author{Jan Kowollik \\
  University of Duisburg-Essen \\
  {\tt jan.kowollik@stud.uni-due.de} \\\And
  Ahmet Aker \\
  University of Duisburg-Essen\\
  {\tt a.aker@is.inf.uni-due.de} \\}

\date{}

\begin{document}
\maketitle
\begin{abstract}
  In this paper we present our system for the FEVER Challenge.
  The task of this challenge is to verify claims by extracting information from Wikipedia.
Our system has two parts. In the first part it performs a search for candidate sentences by treating the claims as query. In the second part it filters out noise from these candidates and uses the remaining ones to decide whether they support or refute or entail not enough information to verify the claim.
  We show that this system achieves a FEVER score of 0.3927 on the FEVER shared task development data set which
  is a 25.5\% improvement over the baseline score.
\end{abstract}

\section{Introduction}

In this paper we present our system for the FEVER Challenge\footnote{\url{http://fever.ai}}.
The FEVER Challenge is a shared task on fact extraction and claim verification.
Initially \citet{Thorne18Fever} created an annotated corpus of $185,445$ claims and proposed a baseline
system to predict the correct labels as well as the pieces of evidence for the claims. 

Our system consist of two parts. In the first part we retrieve sentences that are relevant to a claim. The claim is used as query and is submitted to Lucene search API. The sentences found are candidates for pieces of evidence for the claim. Next in the second part we run a modified version of the Decomposable Attention network \citep{parikh2016decomposable} to predict the textual entailment between a claim and the candidate sentences found through searching but also between claim and all candidate sentences merged into one long text. This step gives us entailment probabilities. We also use a point system to filter out some noise (irrelevant sentences). Based on the remaining candidates we perform label prediction, i.e. whether the claim is supported, refuted or there is not enough evidence. Our system achieves a FEVER score of 0.3927 on the FEVER shared task development data set which is a 25.5\% improvement over the baseline score.

\section{Data}

The data consists of two parts: the FEVER data set and the Wikipedia dump \citep{Thorne18Fever}.

The Wikipedia dump contains over five million pages but for each page only the first section was taken.
The text on each page was split into sentences and each sentence was assigned an index.
The page title is written using underscores between the individual words instead of spaces.

The FEVER data set contains the annotated claims that should be correctly predicted by the developed
system. Each claim is annotated with one of the three labels \emph{SUPPORTS} (verifiably true), \emph{REFUTES} (verifiably false) and \emph{NOT ENOUGH INFO} (not verifiable).
For claims with the 
first two labels the pieces of evidence are provided as a combination of Wikipedia page title and sentence index on that page.

The FEVER data set created by \citet{Thorne18Fever} is split into a training set with $145,449$,
a development set with $19,998$ and a test set with $19,998$ annotated claims. 
The development and test sets are balanced while the training set has an approximately
16:6:7 split on the three labels. Each data set and also the Wikipedia dump is available at the
FEVER web page\footnote{\url{http://fever.ai/data.html}}.

\section{System}

We decided to adopt the general two part structure of the baseline for our system with a key difference.
The first part takes the claim and finds candidate sentences that ideally have
a high chance of being evidence for the claim. The second part determines the label and selects evidence sentences. 

The baseline system uses the sentences found in the first part directly as evidence.
In our system we only find candidate sentences in the first part and select the actual evidence sentences at the end of the second part.
This allows us to operate on a larger number of sentences in the second part of the system and achieve higher recall.

\subsection{Finding Candidate Sentences}

The main idea of the first part of our system is to mimic human behavior when verifying a claim.
If we take a claim about a person as an example, a human is likely to just take few keywords such as the person's name and use this to search for the right Wikipedia page to find evidence. We mimic this behavior by first extracting few keywords from the claim and use them to find candidate sentences in the Wikipedia dump.

\subsubsection*{Extracting Keywords}

We use Named Entity Recognition (NER), Constituency Parsing and Dependency Parsing to extract keywords from each claim.
For NER we use the neural network model created by \citet{ELMO2018andNER}. We use all found named entities as keywords.
For the Constituency Parsing we use the neural network model created by \citet{constituencyparser2017}. We extract all NP tagged phrases from the first two 
recursion layers as keywords because we found that this finds mostly subjects and objects of a sentence.
These two neural networks both use the AllenNLP library \citep{allennlp}.
For the dependency parsing we use the Standford Dependency Parser \citep{dependency_parser}.
We extract all subject and object phrases as keywords.

The NE recognition is our main source for keywords extraction while the other two systems provide additional
keywords that either have not been found by the NER or that are not named entities
in the first place. Example of the keywords being extracted from claims shown in Table \ref{tab:claims} are shown in Table \ref{tab:keywords}. 

\begin{table*}[t!]
\centering
\begin{tabular}{llll}
  \# & Named Entity Recognition & Constituency Parser & Dependency Parser \\
  \hline
  1 & Northern Isles, Scotland & The Northern Isles & The Northern Isles \\
  2 & - & Artificial intelligence, concern & Artificial intelligence, concern \\
  3 & Walk of Life & album, the highest grossing album & -\\
\end{tabular}
\caption{Generated keywords from the three systems (see Table \ref{tab:claims} for claims). For the first claim the NE recognition correctly finds
the two named entities while the other two systems miss one entity and got an additional \emph{The} into the keyword. The second claim has no named entities and the other systems correctly find the relevant parts. In the third example the named entity found by the NE recognition is disambiguated by the Constituency Parser.
  }\label{tab:keywords}
\end{table*}

\begin{table}
\centering
\small
\begin{tabular}{|l|l|}
\hline
{\bf \#} & {\bf Claim}\\\hline
1 & The Northern Isles belong to Scotland. \\
2 & Artificial intelligence raises concern. \\
3 & Walk of Life (album) is the highest grossing album.\\\hline
\end{tabular}
\caption{Example claims used in tables \ref{tab:keywords}/\ref{tab:query}.}\label{tab:claims}
\end{table}

\subsubsection*{Indexing the Wikipedia Dump}

After extracting the keywords we use the Lucene search API\footnote{\url{https://lucene.apache.org/core/}}
to find candidate sentences for each claim. Before searching with Lucene the Wikipedia texts need to be indexed. We treat each sentence as a separate document and index it. We exclude sentences that are empty and also those that are longer than 2000 characters.

For each sentence we also add the Wikipedia page title and make it searchable. For the title we replace all underscores with spaces to improve matching.
In each sentence we replace the words \emph{He}, \emph{She}, \emph{It} and \emph{They} with the
Wikipedia page title that the sentence was found in. When looking at an entire Wikipedia page it is
obvious who or what these words refer to but when searching individual sentences we do not have the
necessary context available. We perform this replacement to provide more context.

\subsubsection*{Searching for the Candidate Sentences}

\begin{table*}[t!]
\centering
\begin{tabular}{llll}
  Query type & Query & Occurrence & Limit \\
  \hline
  Type 1 & "Artificial" "intelligence" & must occur & 2 \\
  Type 1 & "concern" & must occur & 2 \\
  Type 2 & "Artificial" "intelligence" "concern" & should occur & 20 \\
  Type 3 & "Artificial intelligence" "concern" & should occur & 20\\
\end{tabular}
\caption{Generated queries for claim 2 (see Table \ref{tab:claims}). Claim 2 has two keywords where one contains two words. For the Type 1 query we create two queries where one query contains two separate words. For the Type 2 query we split all words and use them all in one query. For type 3 we omit the split and use entire keyword phrases as query.
  }\label{tab:query}
\end{table*}

We use three types of queries to search for candidate sentences for a claim:
\begin{itemize}
\item \textbf{Type 1:} For each keyword we split the keyword phrase into individual words and create a query that searches within the Wikipedia page titles requiring all the individual words to be found.
\item \textbf{Type 2:} We split all keywords into individual words and combine those into one query searching within the Wikipedia page titles to find sentences on those pages where as many words as possible match the title of the page.
\item \textbf{Type 3:} We combine all keywords as phrases into one query searching within the sentences to find those sentences where as many keywords as possible match.
\end{itemize}

We limit the number of results to the two most relevant sentences for the first query type and 20 sentences for
the other two queries because the first query type makes one query per keyword while the other two
only make one query per claim. An example of the queries being generated is given in Table \ref{tab:query}.
If the same sentence is found twice we do not add it to the candidate list again.
For each of the candidate sentences we add the Wikipedia page title at the beginning of the sentence
if it does not already contain it somewhere.

\begin{table}
\centering
\small
\begin{tabular}{|l|lll|}
\hline
 & {\bf SUP} & {\bf REF} & {\bf NEI} \\\hline
SUP & 3291 & 370 & 3005 \\
REF & 1000 & 3159 & 2507 \\
NEI & 1710 & 1142 & 3814\\\hline
\end{tabular}
\caption{Confusion matrix of the full system prediction. Columns are predictions and rows the true labels.}\label{tab:withMerge}
\end{table}

\begin{table}
\centering
\small
\begin{tabular}{|l|lll|}
\hline
 & {\bf SUP} & {\bf REF} & {\bf NEI} \\\hline
SUP & -8 & +52 & -44 \\
REF & -2 & +926 & -924 \\
NEI & -4 & +152 & -148 \\\hline
\end{tabular}
\caption{Confusion matrix change due to including the merge feature. Columns are predictions and rows the true labels.}\label{tab:withMergeChange}
\end{table}

\subsection{Making the Prediction}

The second part of our system first processes the candidate sentences in three independent steps that can be run
in parallel:
\begin{itemize}
\item We use a modified version of the Decomposable Attention neural network \citep{parikh2016decomposable}
to predict the textual entailment between each candidate sentence and its corresponding claim.
\item We merge all candidate sentences of a claim into one block of
text and predict the textual entailment between this block of text and the claim.
\item We assign points to each candidate sentence based on POS-Tags. 
\end{itemize}

Finally our system combines the results in order to decide on the label and to predict the evidence sentences for each claim.

\subsubsection*{Textual Entailment}

We started with the Decomposable Attention network \citep{parikh2016decomposable} that is also used in the baseline except that
we predict the textual entailment for each pair of candidate sentence and claim.
We found that for long sentences the network
has high attention in different parts of the sentence
that semantically belong to different statements.
Using the idea that long sentences often contain multiple statements we made the following additions
to the Decomposable Attention network.

We 
include an additional 2-dimensional convolution layer that operates on
the attention matrix in case of sufficiently large sentences.
Based on our testing we decided on a single convolution layer with a kernel size of 12.
The output of this convolution layer contains a different amount of elements depending on the size
of the attention matrix. This makes sense as longer sentences can contain multiple statements.
We use a \emph{CnnEncoder}\footnote{\url{https://allenai.github.io/allennlp-docs/api/allennlp.modules.seq2vec_encoders.html}} to change the different length output into a same length output.
This is necessary in order to use the result of the convolution layer in a later step of the network
and can be seen as a selection of the correct statement from the available data.
The output of the \emph{CnnEncoder} is concatenated to the input of the aggregate step of the network.
If either the claim or the candidate sentence are shorter than 12 then we skip this additional step and concatenate
a zero vector instead.

When predicting the textual entailment we do not reduce the probabilities to a final label immediately
but keep working with the probabilities in the final prediction (see Section Final Prediction).

\subsubsection*{Merge Sentences}

For each claim we merge all the candidate sentences into one block of text similarly to the baseline.
We predict the textual entailment using our modified decomposable attention network.
We found that the \emph{REFUTES} label is predicted with very high accuracy. However, this is not the case for the other two labels.
By including the results of this step we can improve the predicted labels for the \emph{REFUTES} label as shown in Table \ref{tab:withMergeChange}. Comparing that to the full result given in Table \ref{tab:withMerge} we can see that about $29.3\%$ of correct \emph{REFUTES} predictions are due to this step.

\subsubsection*{Creating POS-Tags and Assigning Points}
We use the Stanford POS-Tagger \citep{pos_tagger} to create POS-Tags for all candidate sentences and all claims.
We found that the Stanford POS-Tagger only uses a single CPU core on our system so we wrote
a script that splits the file containing all claim or candidate sentences into multiple files.
Then the script calls multiple POS-Tagger instances in parallel, one for each file.
The results are then merged back into a single file.

Using the generated POS-Tags we assign scores to the candidate sentences. First each candidate sentence is assigned 5 different scores, one for each of the following POS-Tag categories: verbs, nouns, adjectives, adverbs and numbers.
Each category score starts at 3 and is decreased by 1 for each word of the respective POS-Tag category that is in the claim but not in the candidate sentence.
Duplicate words are considered only once.
We do not allow the category scores to go negative.
At the end the category scores are added together to create the final score which can be a maximum of 15.

\begin{figure}
\includegraphics[width=\linewidth]{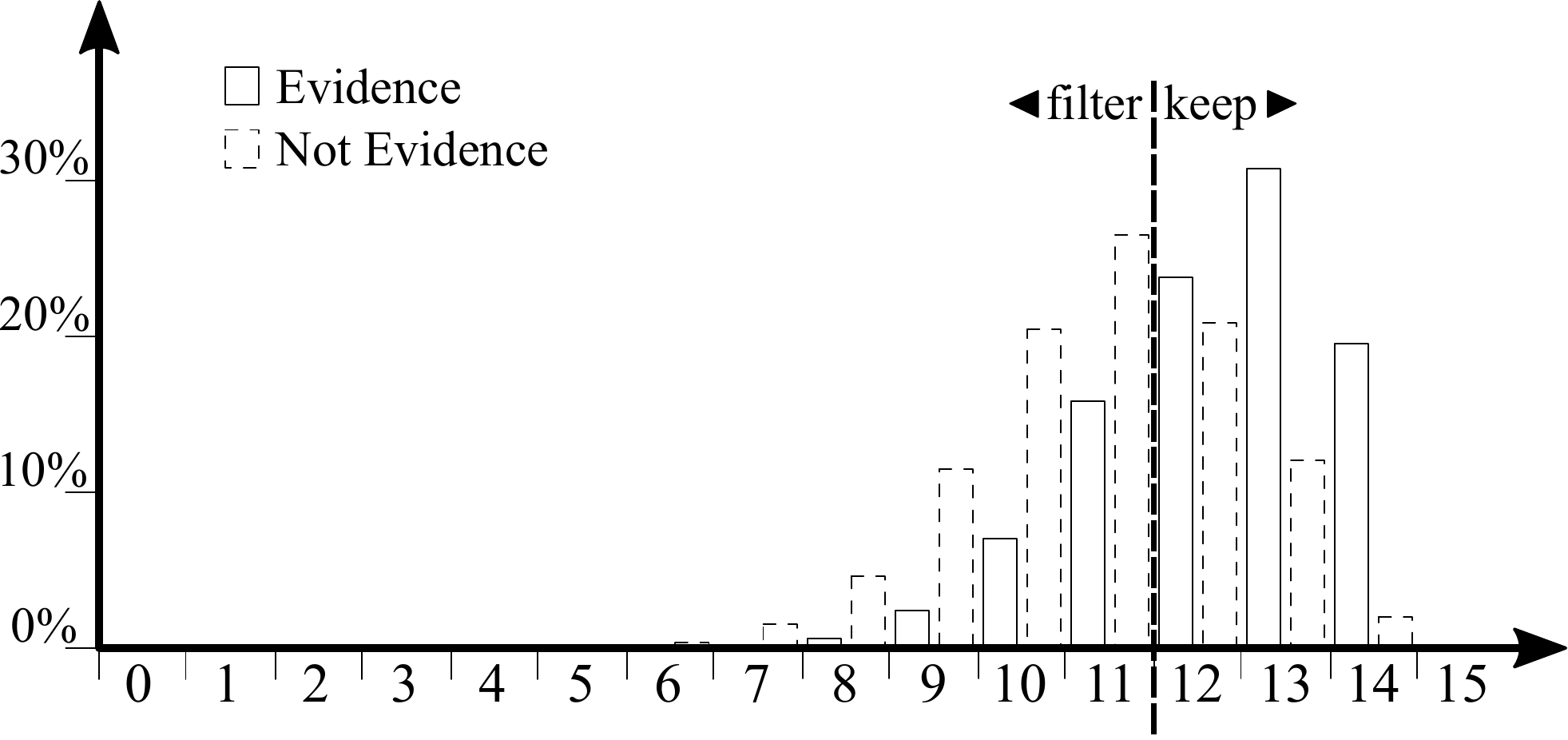}
\caption{Histogram of how many candidate sentences (y-axis) received how many points (x-axis) for the development set. $65.07\%$ of non-evidence and $26.21\%$ of evidence sentences get filtered with the threshold between 11 and 12.
  }\label{tab:point}
\end{figure}

\subsubsection*{Final Prediction}
\label{final_prediction}

We create a matrix from the per candidate sentence textual entailment probabilities with the three labels as columns
and one row per candidate. We reduce all three probabilities of a candidate sentence if it received 11 or less points. The number 11 is empirically determined using the development set. As shown in Figure \ref{tab:point} we are able to filter out most of the non-evidence sentences by looking only at candidate sentences whose point score is more than 11.
Reducing the probabilities is done by multiplying them with $0.3$. This way they are always reduced below the minimum highest probability of non-filtered sentences (= $33.33...\%$).

Finally we predict the label and decide on the evidence sentences.
If the \emph{Merge Sentences} prediction predicted \emph{REFUTES} then we use \emph{REFUTES} as final label.
Otherwise we find the highest value in the matrix and select the column it appears in as final label.
We sort the matrix based on the column of the final label and select the top 5 candidate sentences as evidence.

\subsection{Training}

For training the modified Decomposable Attention network we are using the SNLI data set 
and the FEVER training set \citep{snli,Thorne18Fever}.
For claims labeled as \emph{NOT ENOUGH INFO} 
we first search for Wikipedia page titles
that contain a word from the claim and then randomly choose one of the sentences on that page.
If no Wikipedia page is found this way we randomly select one.
We concatenate the generated training data with the SNLI data set to create
the final training data containing $849,426$ claims.

\section{Results}

Our system achieves a FEVER score of 0.3927 on the shared task development set containing $19,998$ claims.
This is a 25.5\% improvement over the baseline score of 0.3127 on the development set.
The confusion matrix for the predicted labels is given in Table \ref{tab:withMerge}.
It shows that the highest incorrect predictions are for the \emph{NOT ENOUGH INFO} label while the \emph{REFUTES} label is predicted with the least amount of errors.

For the test set our system generated $773,862$ pairs of candidate sentences and claim sentences.
Only for a single claim out of all $19,998$ claims no candidate sentences were found.

For the development set the candidate sentences found in the first part of our system include the actual evidence of 77.83\% of the claims. In comparison the baseline \citep{Thorne18Fever} only finds 44.22\% of the evidence. Our system finds 38.7 sentences per claim on average, while the baseline is limited to 5 sentences per claim.

When looking at how much each feature improves the final score in Table \ref{tab:resultComp},
we can see that the point system using POS-Tags results in the biggest improvement. 

\begin{table}
\centering
\small
\begin{tabular}{|l|lll|}
\hline
 & {\bf Label} & {\bf Recall} & {\bf Score} \\\hline
All & 0.5132 & 0.3581 & 0.3927 \\
Unmodified DA & 0.5170 & 0.3880 & 0.3909 \\
Without Points & 0.4545 & 0.1169 & 0.3665 \\
Without Merge & 0.4747 & 0.3294 & 0.3815 \\
\hline
\end{tabular}
\caption{Contribution of each feature. \emph{Label} refers to the label accuracy, while \emph{Recall} refers to the evidence recall.}\label{tab:resultComp}
\end{table}

\section{Conclusion}

In this paper we have presented our system for the FEVER Challenge.
While keeping the two-part structure of the baseline we replaced the first
part completely and heavily modified the second part to achieve a 25.5\% FEVER score improvement over the baseline. In our immediate future work we will investigate alternative ways of obtaining higher recall in the first part but also improve the textual entailment to further reduce noise.

\bibliography{emnlp2018}
\bibliographystyle{acl_natbib_nourl}

\appendix

\end{document}